\documentclass[%
 aps,
 prmaterials,
 amsmath,amssymb,
reprint,%
longbibliography,
floatfix
]{revtex4-2}

\usepackage{graphicx}
\usepackage{dcolumn}
\usepackage{bm}
\usepackage{amsmath}
\usepackage{multirow}
\usepackage{color}
\usepackage[normalem]{ulem} 

\usepackage{siunitx}
\usepackage[ddmmyyyy,hhmmss]{datetime}

\begin{document}


\title[Lynch {\it et al.} Rydberg Excitons in Synthetic Cuprous Oxide]{Rydberg Excitons in Synthetic Cuprous Oxide (Cu$_2$O)}

\author{Stephen A. Lynch}
 \email{LynchSA@cardiff.ac.uk}
\affiliation{%
School of Physics and Astronomy, Cardiff University, Cardiff CF24 3AA, United Kingdom.
}%

\author{Ravi P. Singh}
\affiliation{%
Department of Physics, \mbox{Indian Institute of Science Education and Research (IISER) Bhopal, India.}
}%

\author{Chris Hodges}
\affiliation{%
School of Physics and Astronomy, Cardiff University, Cardiff CF24 3AA, United Kingdom.
}%

\author{Soumen Mandal}
\affiliation{%
School of Physics and Astronomy, Cardiff University, Cardiff CF24 3AA, United Kingdom.
}%

\author{Wolfgang Langbein}
\affiliation{%
School of Physics and Astronomy, Cardiff University, Cardiff CF24 3AA, United Kingdom.
}%

\author{\mbox{Liam A. P. Gallagher}}
\affiliation{%
Joint Quantum Centre Durham-Newcastle, Department of Physics, Durham University, Durham DH1 3LE, United Kingdom.
}%
\author{\mbox{Jon D. Pritchett}}
\affiliation{%
Joint Quantum Centre Durham-Newcastle, Department of Physics, Durham University, Durham DH1 3LE, United Kingdom.
}%
\author{Danielle Pizzey}
\affiliation{%
Joint Quantum Centre Durham-Newcastle, Department of Physics, Durham University, Durham DH1 3LE, United Kingdom.
}%

\author{\mbox{Joshua P. Rogers}}
\affiliation{%
Joint Quantum Centre Durham-Newcastle, Department of Physics, Durham University, Durham DH1 3LE, United Kingdom.
}%

\author{Charles S. Adams}
\affiliation{%
Joint Quantum Centre Durham-Newcastle, Department of Physics, Durham University, Durham DH1 3LE, United Kingdom.
}%

\author{Matthew P. A. Jones}
\affiliation{%
Joint Quantum Centre Durham-Newcastle, Department of Physics, Durham University, Durham DH1 3LE, United Kingdom.
}%

\date{\today~at~\currenttime}

\begin{abstract}
High-lying Rydberg states of Mott-Wannier excitons are receiving considerable interest due to the possibility of adding long-range interactions to the physics of exciton-polaritons. Here, we study Rydberg excitation in bulk synthetic cuprous oxide grown by the optical float zone technique and compare the result with natural samples.  X-ray characterization confirms both materials are mostly single crystal, and  mid-infrared transmission spectroscopy revealed little difference between synthetic and natural material. The synthetic samples show principal quantum numbers up to  $n=10$, exhibit additional absorption lines, plus enhanced spatial broadening and spatial inhomogeneity. Room temperature and cryogenic  photoluminescence measurements reveal a significant excess of copper vacancies in the synthetic material. These measurements provide a route towards achieving \mbox{high-$n$} excitons in synthetic crystals, opening a route to scalable quantum devices.

%
\end{abstract}

\maketitle

\section{Introduction}

The emerging research field of Rydberg excitons in cuprous oxide is facilitating an exciting crossover between two previously siloed disciplines: cold-atom physics and semiconductor quantum optics. The coherent spectroscopy of Rydberg states in gas-phase atomic systems has garnered great recent interest, with applications in the measurement of electromagnetic fields from DC to terahertz frequencies \cite{sibalic18,meyer20}, quantum simulation \cite{browaeys20} and computation \cite{saffman16}, and quantum optics \cite{firstenberg16}. There is clear interest in observing similar states in semiconductor systems, where half a century of development in state-of-the-art nano-fabrication techniques provide capabilities that go way beyond those available in the gas phase.  Cuprous oxide raises the tantalizing prospect of a material where Rydberg atom analogues can be created on-demand in an environment that more naturally lends itself to engineering multiple quantum interactions, and furthermore offers a plausible pathway to scale up. Experiments in high-quality natural crystals have revealed the potential of cuprous oxide but realizing synthetic material with comparable quantum optical properties is a prerequisite for the material to achieve technological traction. In this paper we discuss our progress towards achieving synthetic cuprous oxide single crystals that exhibit \mbox{high-$n$} excitons.

Single crystal cuprous oxide (Cu$_2$O) was one of the first semiconducting crystals discovered~\cite{grondahl27,grondahl33}, and received considerable attention until silicon eventually prevailed as the semiconductor of choice for microelectronic applications. However, it is the unique optical properties that cause it to stand out from other semiconductor crystals. The Cu$_2$O band gap of 2.17~eV is about twice that of silicon, making it suited to complement silicon as a solar cell material~\cite{chen13}. The spectroscopic property of paramount interest to us is its extraordinary excitonic spectrum. Excitons are bound electron and hole quasi-particles that exhibit a series of sharp spectral lines (not unlike the Rydberg series of hydrogen) near the semiconductor band edge. Because the exciton binding energy is typically in the meV range, the exciton lines are mostly observed at cryogenic temperatures. While it is common to observe excitons up to principal quantum numbers of $n=3$ in conventional compound semiconductors such as GaAs~\cite{weisbuch00}, it has long been known that excitons with \mbox{higher-$n$} can be observed in Cu$_2$O crystals~\cite{baumeister61}. The Cu$_2$O-exciton system has been used over the last three decades to test for evidence of Bose-Einstein condensation in the solid state~\cite{snoke14}, due to its dipole forbidden exciton ground state transition.

However, it is the 2014 paper by Kazimierczuk {\it et al.} reporting excitons with $n=25$ that has generated renewed interest in this material system~\cite{kazimierczuk14}. Excitons may be described using a modified Bohr model~\cite{dexter65,fox03}, with the energy level spectrum described by,
\begin{equation}\label{eqn:energy}
E_{\mathrm{X}}(n) = -\left(\dfrac{\mu}{m_0}\right)\dfrac{1}{\epsilon_r^2}\dfrac{R_{\mathrm{H}}}{n^2} = -\dfrac{R_{\mathrm{X}}}{n^2} \enskip,
\end{equation}
and exciton radius,
\begin{equation}
\left\langle r_n \right\rangle = n^2 \left(\dfrac{m_0}{\mu}\right) \epsilon_r a_\mathrm{H} = n^2 a_{\mathrm{X}} \enskip.
\end{equation}
Here $\mu$ is the two-body reduced mass, $R_{\mathrm{H}}$ and $R_{\mathrm{X}}$ are the hydrogen and exciton Rydberg constants respectively, $a_\mathrm{H}$ and $a_\mathrm{X}$ are the hydrogen and exciton Bohr radii respectively, and the other symbols have their usual meanings. For the results in~\cite{kazimierczuk14}, the relevant excitonic states give rise to the ``yellow'' series of transitions between the upper and lower states of the spin-orbit split conduction and valence bands. Since the band states have the same parity, the excitonic energy levels have P (i.e. orbital angular momentum $l=1$) symmetry. A narrow quadrupole transition to the dipole-forbidden 1S state is also be observed, and phonon-assisted 1S transitions contribute a substantial background to the spectrum of nP states. The electron and hole effective masses are usually taken to be~\cite{hodby76} \mbox{$m_e=0.99~m_0$} and \mbox{$m_{h}=0.69~m_0$}, which gives $\mu = 0.41~m_0$. There is some discussion over the correct value of $\epsilon_r$ to use. The high-frequency relative permittivity value is often quoted as~\cite{hodby76} $\epsilon_r(\infty) = 6.46$, which gives \mbox{$R_{\mathrm{X}} = 132.6$~meV}. However, since the ground state 1S exciton is known to have an anomalously large binding energy~\cite{kavoulakis97}, this value is not used. Instead, researchers use the modified Bohr formula (equation~\ref{eqn:energy}) to fit the energies of the higher lying P-excitons, yielding a smaller exciton Rydberg energy of \mbox{98.3~meV}, which corresponds to $\epsilon_r = 7.5$. In reality the situation is more complicated. When the non-parabolicity of the bands, phonon coupling, and the spin interaction are properly taken into account~\cite{kavoulakis97}, the adjusted Bohr radius of the exciton is calculated to be \mbox{$a_\mathrm{X} = 1.11$~nm}. Kazimierczuk {\it et al.} use this value together with an average radius $\langle r_n \rangle$ derived from the spherical harmonics of the $n = 25$ P-state wave-functions to predict $\langle r_{25} \rangle = 1.04$~\si{\micro\meter}. In atomic systems, the large spatial extent of excitonic wave-functions means that interactions between them are dominated by long-range van der Waals forces~\cite{walther18}. Unlike the short-range exchange-type interaction more commonly studied for ground state excitons, such long-range interactions can enable strong non-local~\cite{busche17} optical non-linearities~\cite{lukin01,pritchard10} without overlap between the excitonic wave-functions. While widely studied in atomic systems~\cite{firstenberg16}, the first hints of such long-range interactions in crystals were observed in Kazimierczuk {\it et al.}~\cite{kazimierczuk14}, with subsequent confirmation via pump-probe experiments~\cite{heckotter18}. 

So far however, these results~\cite{kazimierczuk14} have not been reproduced by other groups, due to difficulties in obtaining samples of similar quality. Most groups currently report on gemstones of various quality (Fig.~\ref{fig:cuprite_images}(a)). The situation is particularly acute for synthetic samples of cuprous oxide, where the highest principal quantum number measured to date is $n=10$~\cite{ito98}. Wider availability of the high-quality samples required for the observation of \mbox{high-$n$} excitons is an essential step towards various quantum-optical devices based on Rydberg excitons~\cite{khazali17,walther18giant,zielinska19,walther20}.

 Several methods for producing Cu$_2$O have been reported in the literature. Thin films have been prepared by several deposition methods, including reactive sputtering~\cite{ogale92}, thermal oxidation~\cite{musa98}, electro-deposition~\cite{santra99,mukhopadhyay92}, melting~\cite{naka05}, and metal-organic chemical vapour deposition (MOCVD)~\cite{jeong09}. However, the experimental conditions necessary to observe \mbox{high-$n$} excitons impose severe constraints on the thermal expansion properties of any supporting substrate to prevent strain accumulation in the thin film on cooling to cryogenic temperatures. This is because the large spatial extent of the \mbox{high-$n$} exciton wave-function should make it particularly sensitive to any distortions to the crystal lattice resulting from macroscopic strain fields~\cite{agekyan77,kruger18}. Consequently, we have chosen to pursue a path towards macroscopic single crystal growth using float-zone (FZ) refining, to avoid the thermal strain that would occur on cryogenic cooling if the Cu$_2$O was bonded to a different substrate. This method has been successfully employed in the past to produce large single crystals~\cite{schmidt_whitley74, ito98}. 

\section{METHODS}

\subsection{Crystal Growth}
The starting material was 5~mm diameter copper rod, Puratronic\textsuperscript{\textregistered} 99.999\% (metals basis) from Alfa Aeser. The Cu$_2$O feed and seed rods were prepared by thermal oxidation of the copper metal rods at 1100$^\circ$C for approximately 40 hours in air. The oxidized rods were then used to grow the single crystal material in a Crystal Systems Corporation (CSC) optical furnace (model: FZ-T-10000-H-VII-VPM-MII-PC). Localized heating was achieved by bringing the light from four halogen lamps to a joint focus on the sample using four semi-ellipsoidal mirrors. The growth region is isolated from the furnace by a quartz tube allowing the use of different growth atmospheres and pressures. Single-crystal growth is achieved by melting a feed rod and fusing it to a seed rod, establishing a freely floating molten zone. This floating zone was then scanned up the feed rod at a controlled slow speed of 4.0-7.0~mm/hour, with the feed and seed rods rotating at 20~rpm in opposite directions. This method has the advantage that it doesn't require a crucible, which is critical in this application because molten Cu$_2$O is known to react with standard crucible materials~\cite{schmidt_whitley74}.
\begin{figure}[!th]
\begin{tabular}{cc}
	\includegraphics[width=4.15cm]{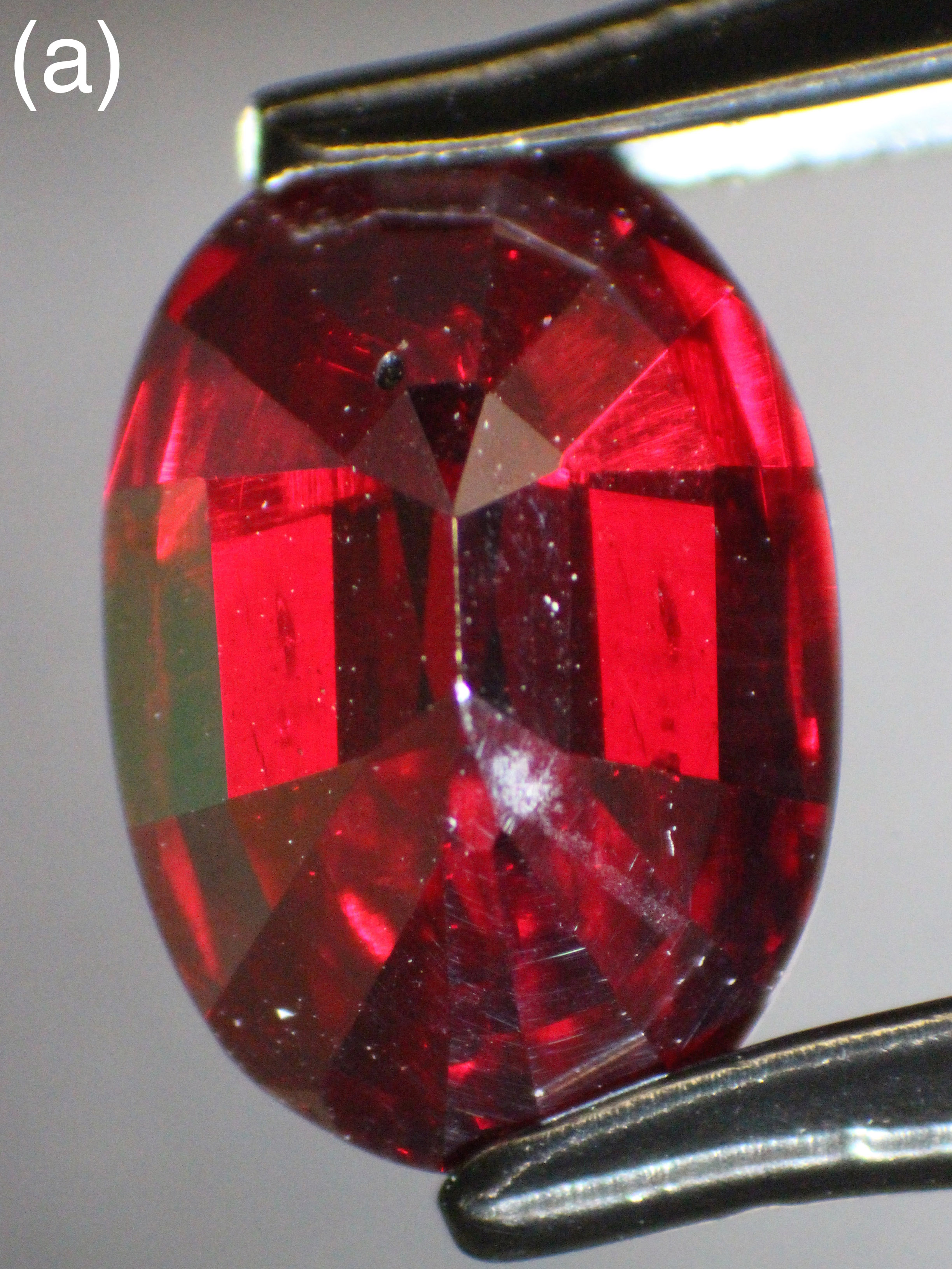} & \includegraphics[width=4.2cm]{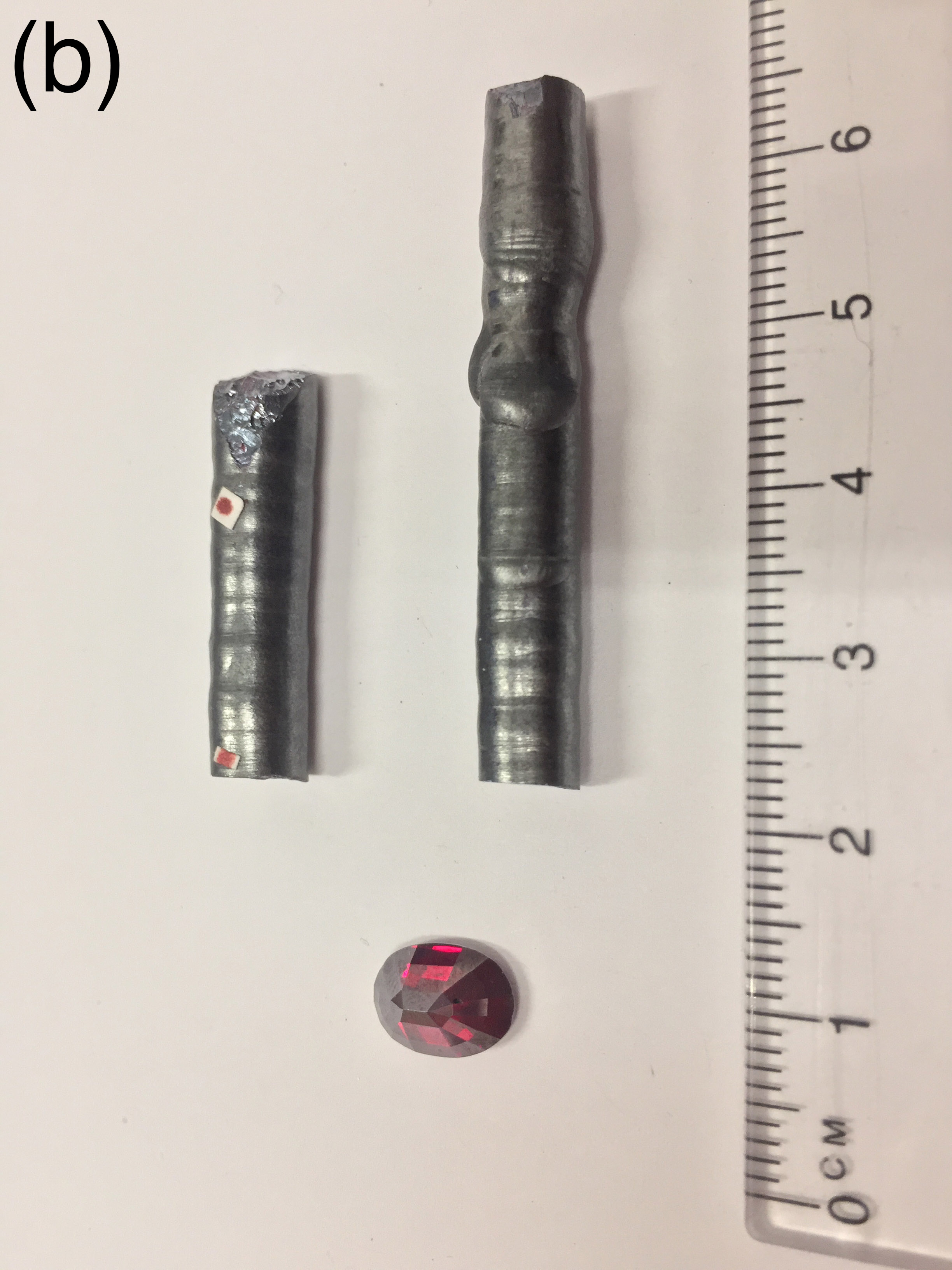} \\
	\includegraphics[width=4.15cm]{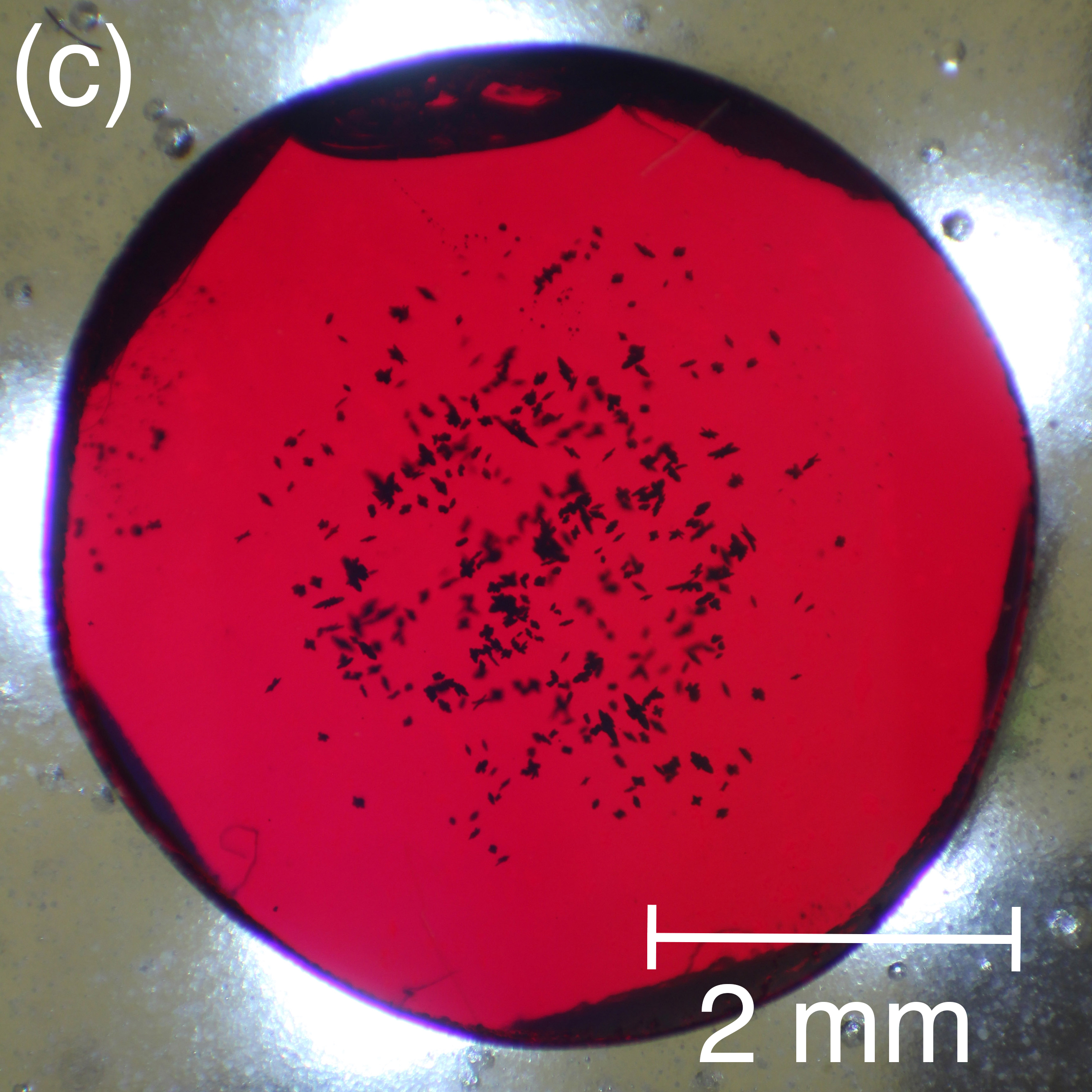} & \includegraphics[width=4.2cm]{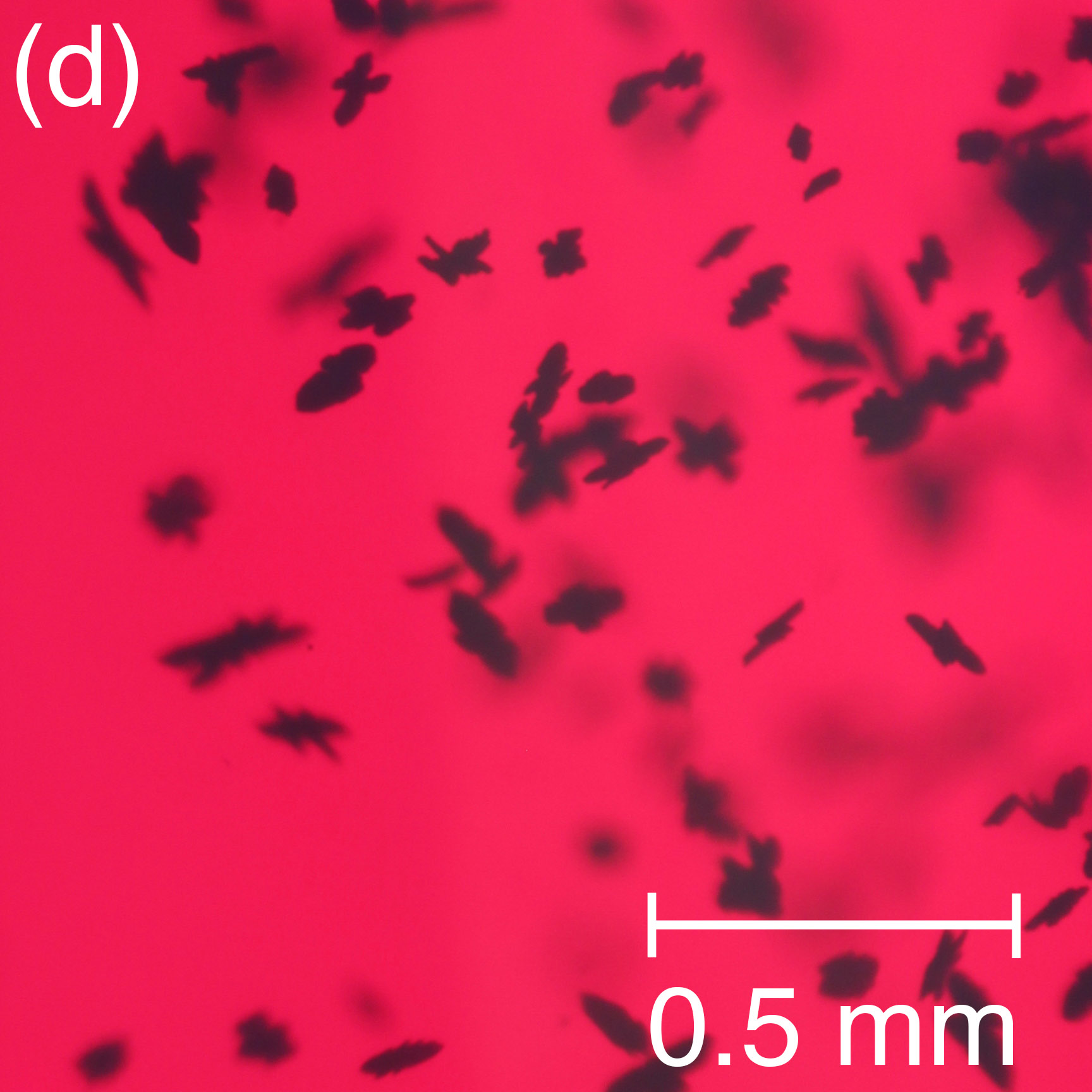}
\end{tabular}
\caption{(a) A natural cuprite gemstone. (b) Two synthetic single crystals of Cu$_2$O with the same gemstone for scale. (c) Back-lit slice of optically polished synthetic crystal. (d) Higher magnification optical image of the same slice highlighting the inclusions (dark specs).}
\label{fig:cuprite_images}
\end{figure}
Figure~\ref{fig:cuprite_images}(b) shows two 5~mm Cu$_2$O single crystals after FZ growth together with the same gemstone shown in (a), with a perspex  lab ruler on the right for scale.

\subsection{Sample Preparation}
For characterization, the crystals were cut into 2-3~mm thick disks perpendicular to the cylindrical axis using a diamond saw. The samples were encapsulated in acrylic resin and lapped back to reveal the front surface of the Cu$_2$O. The uncovered {Cu$_2$O} surface was then polished optically flat on a Struers LaboPol-5 lapping machine, employing a 4-step process using Struers consumables. \mbox{Step~1} was a rough grind of the surface using a MD-Largo disk together with DiaPro Allegro/Largo 9~\si{\micro\meter} diamond suspension/lubricant for 2 minutes. Step 2 was a rough polish on a Struers DP-DAC satin woven acetate cloth disk together with DiaPro DAC 3~\si{\micro\meter} diamond suspension/lubricant for a further 5 minutes. Step 3 was a fine polishing step on a DP-NAP cloth together with a 1~\si{\micro\meter} DiaPro NAP suspension/lubricant for further 15-30 minutes. Step 4 covered the final polishing, where we used a new DP-NAP cloth together with 0.25~\si{\micro\meter} DiaPro NAP suspension/lubricant. During the final polishing step, the sample was periodically inspected under a stereo-microscope, and this step was repeated until visible surface scratches/digs were removed. The other side of the sample was then lapped and polished according to the same procedure. The final thickness of the polished samples was between 50-100~\si{\micro\meter} for optical measurements. Thicker samples (2-3~mm), prepared in a similar way, were used for mid-infrared measurements. A typical low-magnification cross-sectional image is shown in figure~\ref{fig:cuprite_images}(c), showing homogeneous optically transparent material with the deep red signature color of Cu$_2$O near the outside of the disk and an increasing concentration of small dark inclusions localized towards the center. Figure~\ref{fig:cuprite_images}(d) shows the same cross-sectional slice captured at greater magnification, highlighting the inclusions towards the center of the disk. Inclusions have been reported in similar float-zone refined single crystals and were found to be voids with cupric oxide (CuO) walls~\cite{schmidt_whitley74,frazer15iop}.

\subsection{X-ray Analysis}
\begin{figure}[!t]
\includegraphics[width=8.4cm]{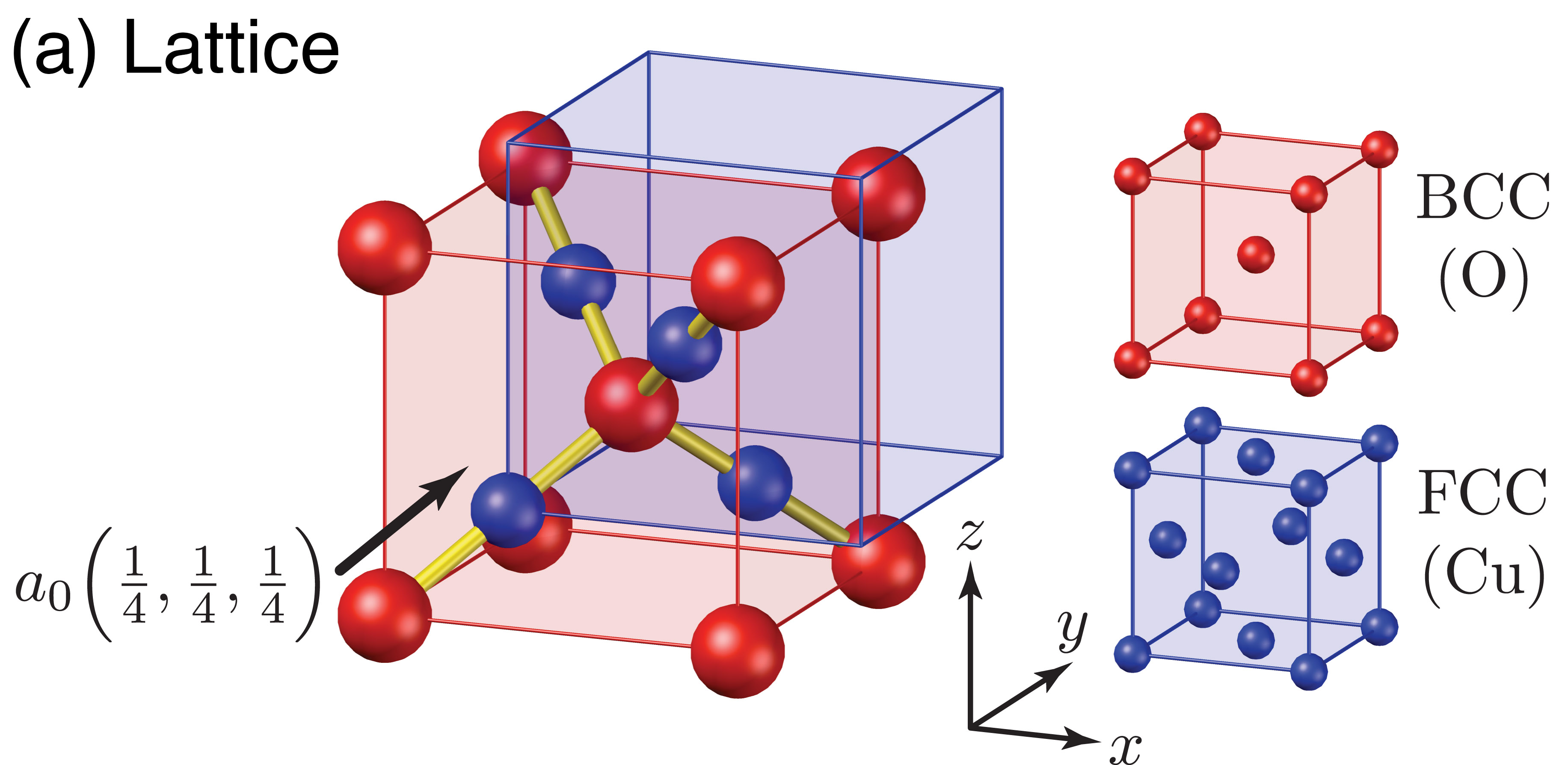}
\begin{tabular}{cc}
\includegraphics[width=4.15cm]{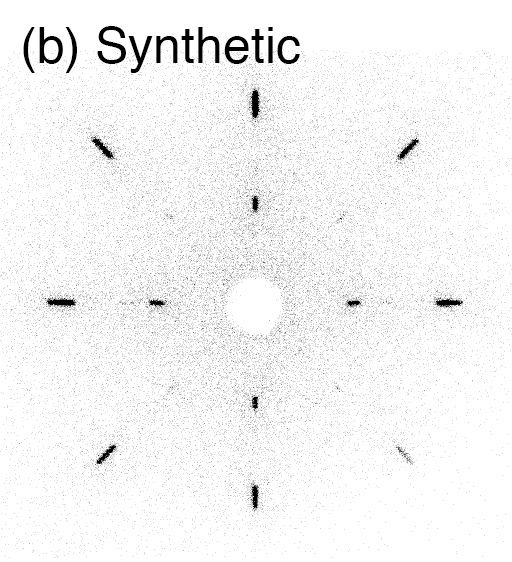} & \includegraphics[width=4.15cm]{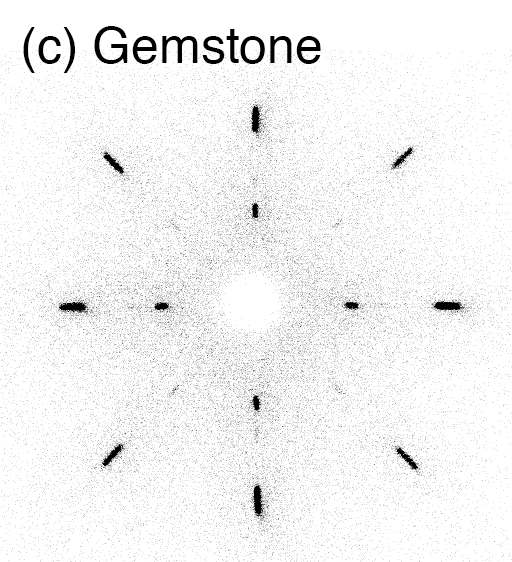} \\
\end{tabular}
\caption{(a) Interleaved BCC/FCC lattice of Cu$_2$O. The oxygen (O) atoms are located at the lattice points of the BCC cubic cell, while the copper (Cu) atoms lie at the lattice points of the FCC cubic cell with the same dimensions ($a_0$). The two cubic cells are offset by $a_0\big(\tfrac{1}{4}, \tfrac{1}{4}, \tfrac{1}{4} \big)$. Laue spot pattern of both the (b) synthetic and (c) natural material for comparison. The crystals were both aligned so that the incoming beam X-ray beam is orthogonal to the 100 plane $\big(1~0~0\big)$ crystal plane. These Laue patterns highlight the cubic symmetry and single crystal nature of both samples of Cu$_2$O material.}
\label{fig:laue_spots}
\end{figure}

\mbox{X-ray} analysis was performed using a Laue camera. The back-scattered Laue spot pattern when the $\big(1~0~0\big)$ crystal plane is aligned perpendicular to the X-ray beam is shown in figure~\ref{fig:laue_spots}(b) and (c) for both synthetic and natural material respectively. 
These Laue patterns confirm the single crystal nature of both the synthetic Cu$_2$O material, and the natural gemstones. Cu$_2$O has an interleaved cubic lattice. The Cu atoms sit at the lattice points of a face-centered cubic (FCC) cell, while the oxygen atoms are located at the lattice points of a shifted body centered cubic (BCC) lattice, as shown in figure~\ref{fig:laue_spots}(a). The expected cubic symmetry of the crystal is clear from the Laue patterns in the bottom of this figure.

\subsection{Visible Spectroscopy}
\begin{figure}[htp]
	\includegraphics[width=8.4cm]{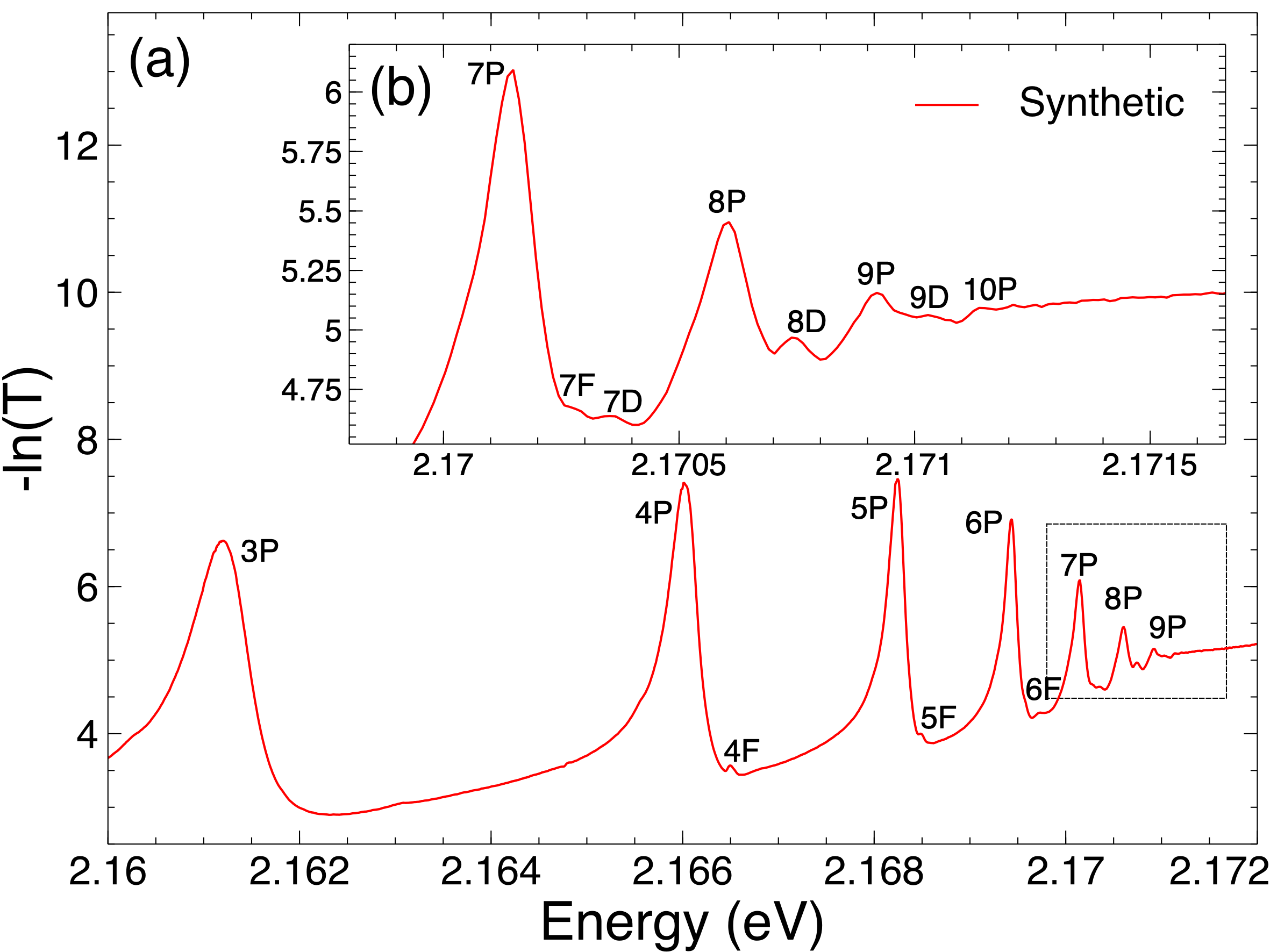} \\
	\includegraphics[width=8.4cm]{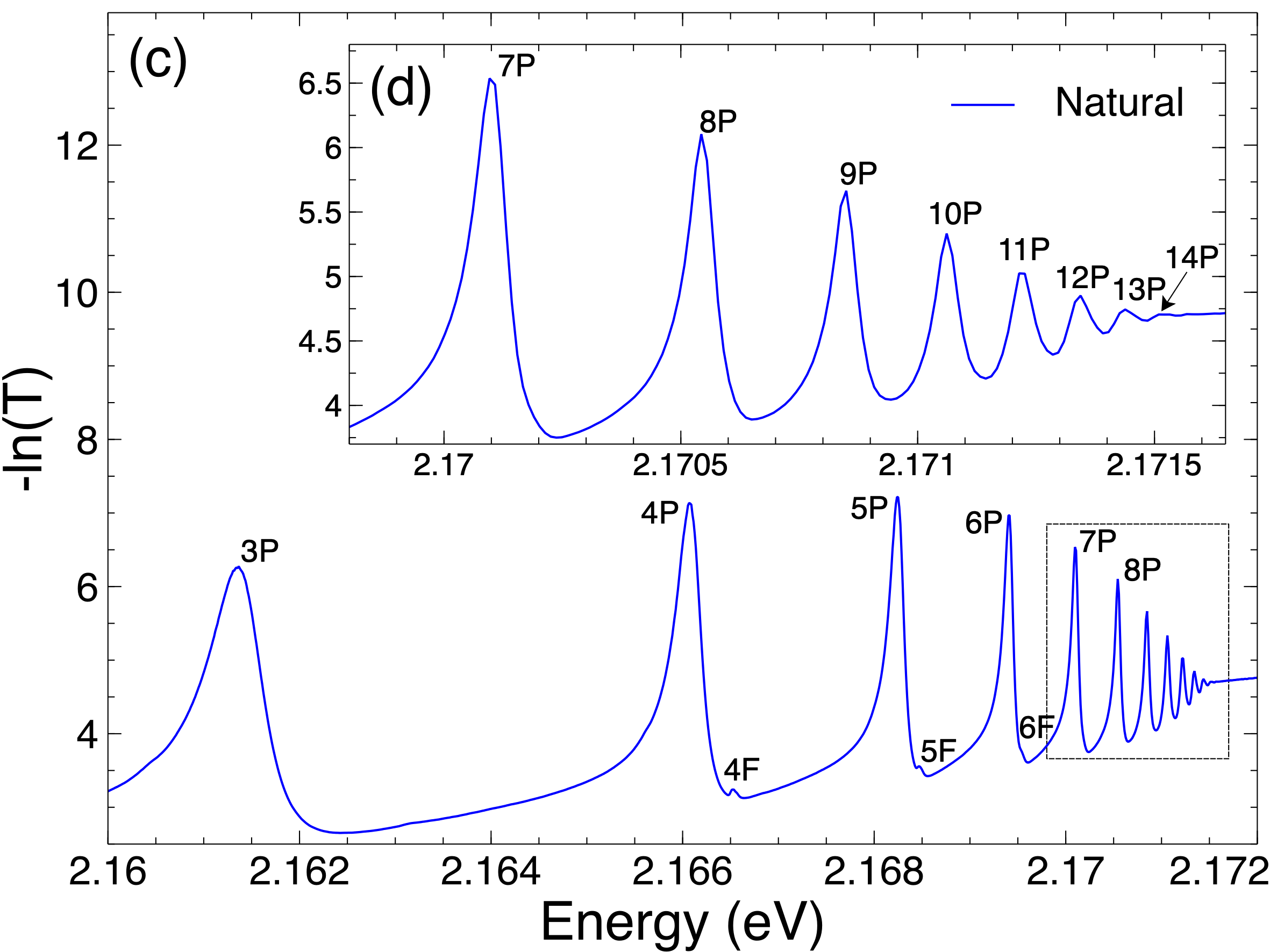} \\
\caption{Transmission spectrum for (a) synthetic (thickness 80~\si{\micro\meter}) and (c) natural (thickness 80~\si{\micro\meter}) samples of Cu$_2$O, measured at a nominal heat-sink temperature of 5~K. Insets (b) and (d) show the \mbox{high-$n$} regions delineated by boxes.
\label{fig:exciton_spectra} }
\end{figure}

The Rydberg exciton spectrum at 5~K for our synthetic material is shown in 
Figure \ref{fig:exciton_spectra}, along with a corresponding spectrum measured for the natural gemstone material for comparison.
The samples were illuminated by a Lumileds LXML-PX02 lime-green light emitting diode with emission wavelength centered at 565~nm, limited to 570-580~nm using an Omega optical 575BP10 band-pass filter, and the exciton spectra were recorded in transmission geometry. The excitation intensity was around 10~mW/cm$^2$. Spectra were also recorded at lower excitation intensities to check that there were no power dependent effects. The spectra were obtained using a custom-made imaging spectrograph with a focal length of 1.9~m. The light was dispersed by a 1200~l/mm holographic grating of ($120 \times 140$)~mm$^2$ size, 900~nm blaze wavelength, and detected by a CCD (Roper Pixis) of $1340 \times 100$ square pixels of 20~\si{\micro\meter} size. The spectromete has a resolution of 30~\si{\micro}eV (full width at half maximum (FWHM)) at 573~nm (2.16~eV) in second order. Sample cooling was provided by a low-vibration closed-cycle cryostat (Montana Cryostation C2) equipped with a XYZ piezo stage (Attocube) that provided a spatial resolution around 0.1~\si{\micro\meter}, allowing focusing and lateral alignment.

A Rydberg series is clearly visible in both cases, extending to $n=9$ (possibly $n=10$) in the synthetic material, and $n=14$ in the natural gemstone sample. These spectra show that our synthetic material is of high-quality, matching the highest principal quantum number previously observed for synthetic crystals~\cite{ito98}. However, exciton peaks above $n=10$ are clearly missing. In both the natural and synthetic stones, from $n=4$, we see small shoulders on the high energy flank of the P states. Due to their energy, and presence only for $n\geq4$, we attribute these peaks to F states~\cite{thewes15}. However, the synthetic sample shows additional peaks between the P and F peaks. By comparing the energy of these peaks to spectra measured through two-photon spectroscopy~\cite{mund18} we conclude these additional peaks are D states. The observation of D states in one-photon measurements is forbidden due to parity. However, if the crystal symmetry is broken, the selection rules are broken, and forbidden angular momentum states can be observed~\cite{heckotter17,heckotter18b}.
\begin{figure}[!htp]
\begin{tabular}{rr}
	\includegraphics[width=7.7cm]{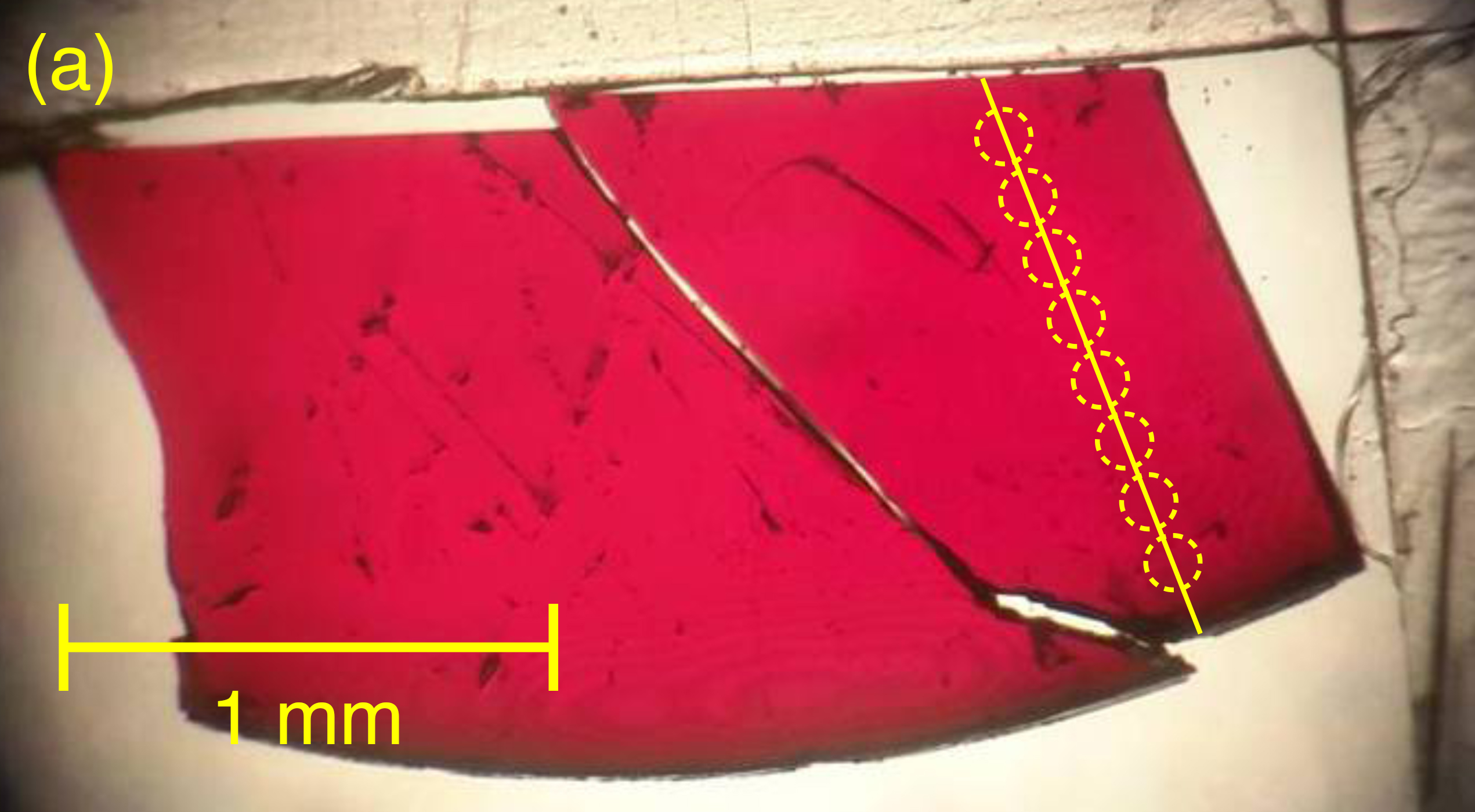} \\
	\includegraphics[width=8.4cm]{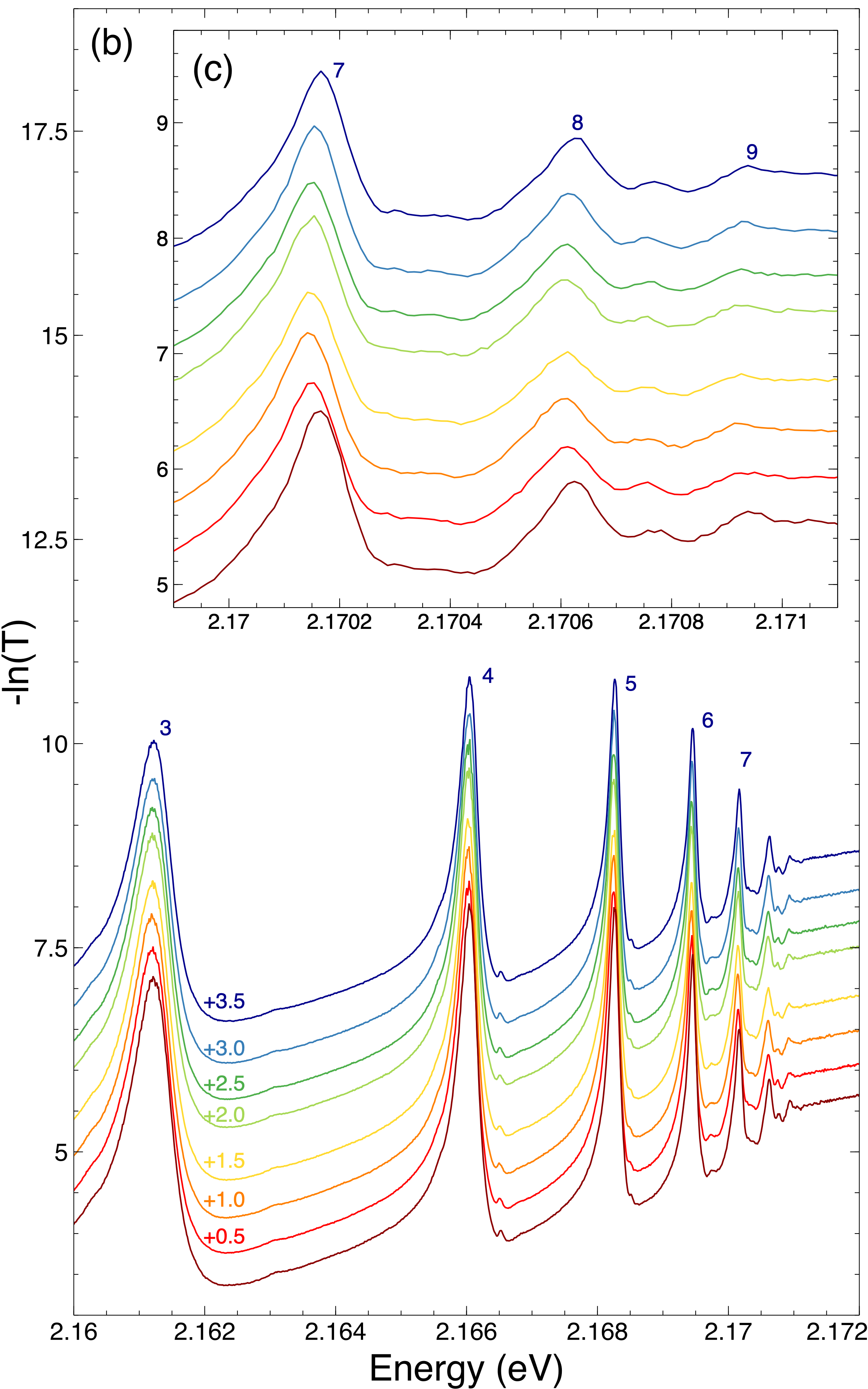}
\end{tabular}
\caption{\label{fig:spatial} (a) Low-magnification optical image of the synthetic Cu$_2$O sample, sandwiched between two CaF$_2$ cover-slips and confined by a spacer of slightly greater thickness than the sample. A small drop of glue that has leaked from the spacer anchors the bottom corner of piece of the sample on the right hand side. The left piece of the sample is mechanically free. (b) Spectral information extracted from different positions along the approximate radial cross-section of the parent rod indicated by the yellow dashed circles. Individual spectra have been offset by multiples of 0.5 to aid comparison. The inset (c) of this figure shows the same data on a magnified energy scale.}
\end{figure}
In a 1977 review paper on the spectroscopy of direct-gap semiconductors~\cite{agekyan77}, possible causes of perturbation of exciton spectra are discussed. The author references an earlier theoretical investigation~\cite{song65}, where it had been revealed that point defects lead to the main line broadening of the exciton line, while large-size defects (dislocations, for example) originate a low-gradient electric field which induces forbidden lines. We conjecture that both effects (the absence of \mbox{high-$n$} excitons and the appearance of the D states) are related, and are due to disorder in the synthetic crystal which breaks the crystal symmetry and affects the overlap integral of the exciton wave-functions. In the remainder of this paper we study two possible sources of such disorder: larger-scale inhomogeneity due to strain, and localized point defects.

\begin{figure}[htp]
	\includegraphics[width=8.4cm]{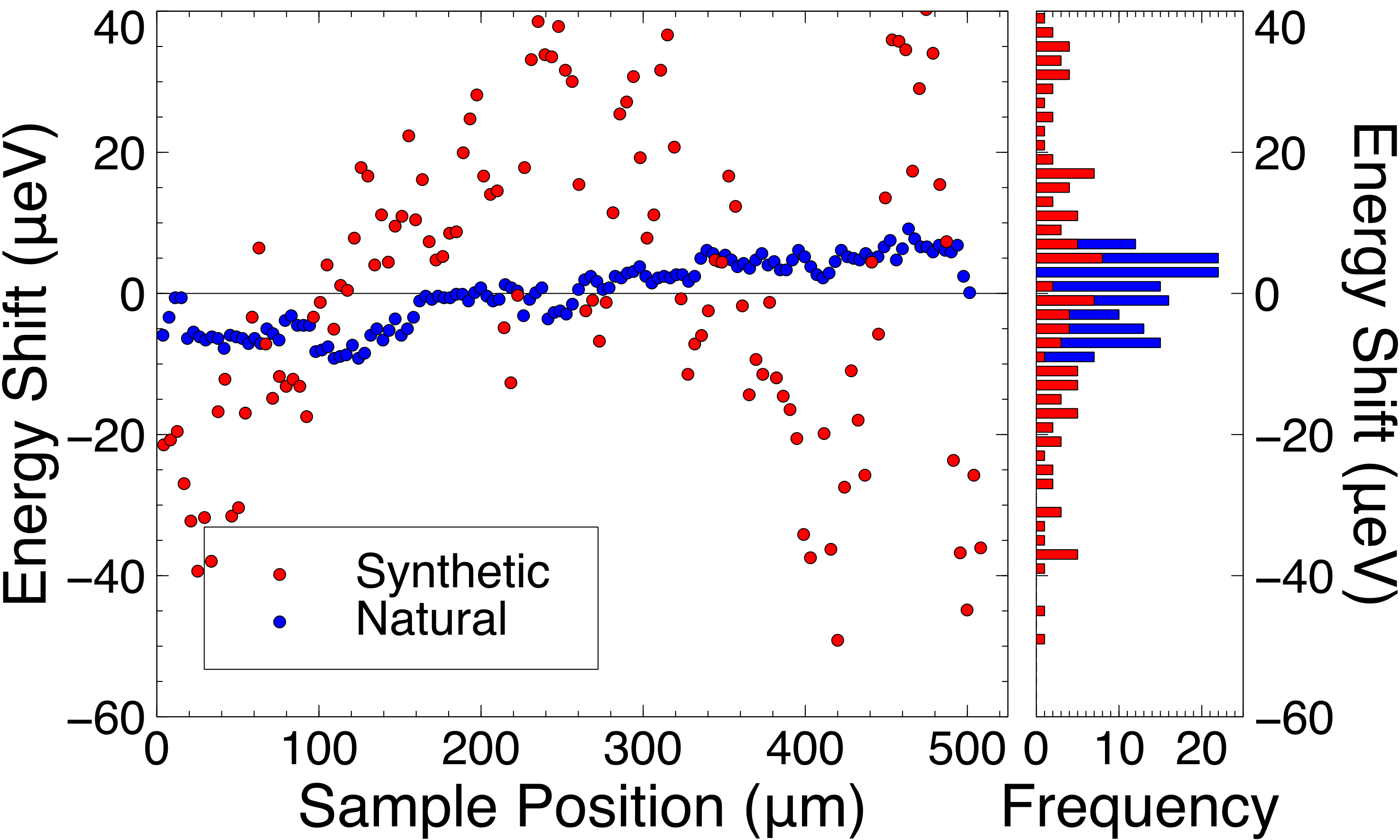}
	\caption{\label{fig:variation_n6} Variation in the energy of the $n=6$ peak for synthetic and natural samples, as a function of position across the sample. Here the spatial resolution was approximately 4.2~\si{\micro\meter} for the synthetic stone and 3.8~\si{\micro\meter} for the natural stone. On the right, histograms show the distribution of energy shift.}
\end{figure}

Strain leads to shifts and splittings of the exciton energy levels~\cite{trebin81}. If the strain is spatially inhomogeneous, then averaging, for example through the thickness of the sample, can cause broadening, with the eventual loss of visibility of \mbox{higher-$n$} peaks.  By fitting the peaks in figure~\ref{fig:exciton_spectra}(a) with an asymmetric Lorentzian profile~\cite{toyozawa58}, we find that the energies of the P states are within 30~\si{\micro\eV} of previously reported values~\cite{kazimierczuk14}.  Figure~\ref{fig:spatial} shows the exciton spectra recorded at different positions along an approximately radial cross-section of the parent rod, at increments of $\sim 100$~\si{\micro\meter}. The overall structure of the exciton spectrum is remarkably uniform, with the additional peaks and the cut-off at \mbox{high-$n$} showing almost no variation across the sample. Close inspection of inset \ref{fig:spatial}(c) reveals slight shifts in the energy of the exciton peaks, which we analyse in more detail in Fig.~\ref{fig:variation_n6}. Here we show the difference between the local energy of the $n=6$ peak $E_6(r)$ and its whole sample mean value $\bar{E}_6$ as a function of position for both the synthetic and natural samples. The dispersion observed for the synthetic sample is significantly larger than for our best natural sample. Here the spread in $E_6(r)$ was 18.3~$\si{\micro\rm{eV}}$ for the natural sample, as compared to 103.9~$\si{\micro\rm{eV}}$ for the synthetic sample, measured over a similar scan range. This indicates that local variation in the strain does occur, but given the similarity between the spectra in Fig.~\ref{fig:spatial}, this does not seem a plausible explanation for the observation of dipole forbidden exciton states and the absence of higher-$n$ states.

\subsection{Photoluminescence Spectroscopy}
\begin{figure}[htp]
\includegraphics[width=8.4cm]{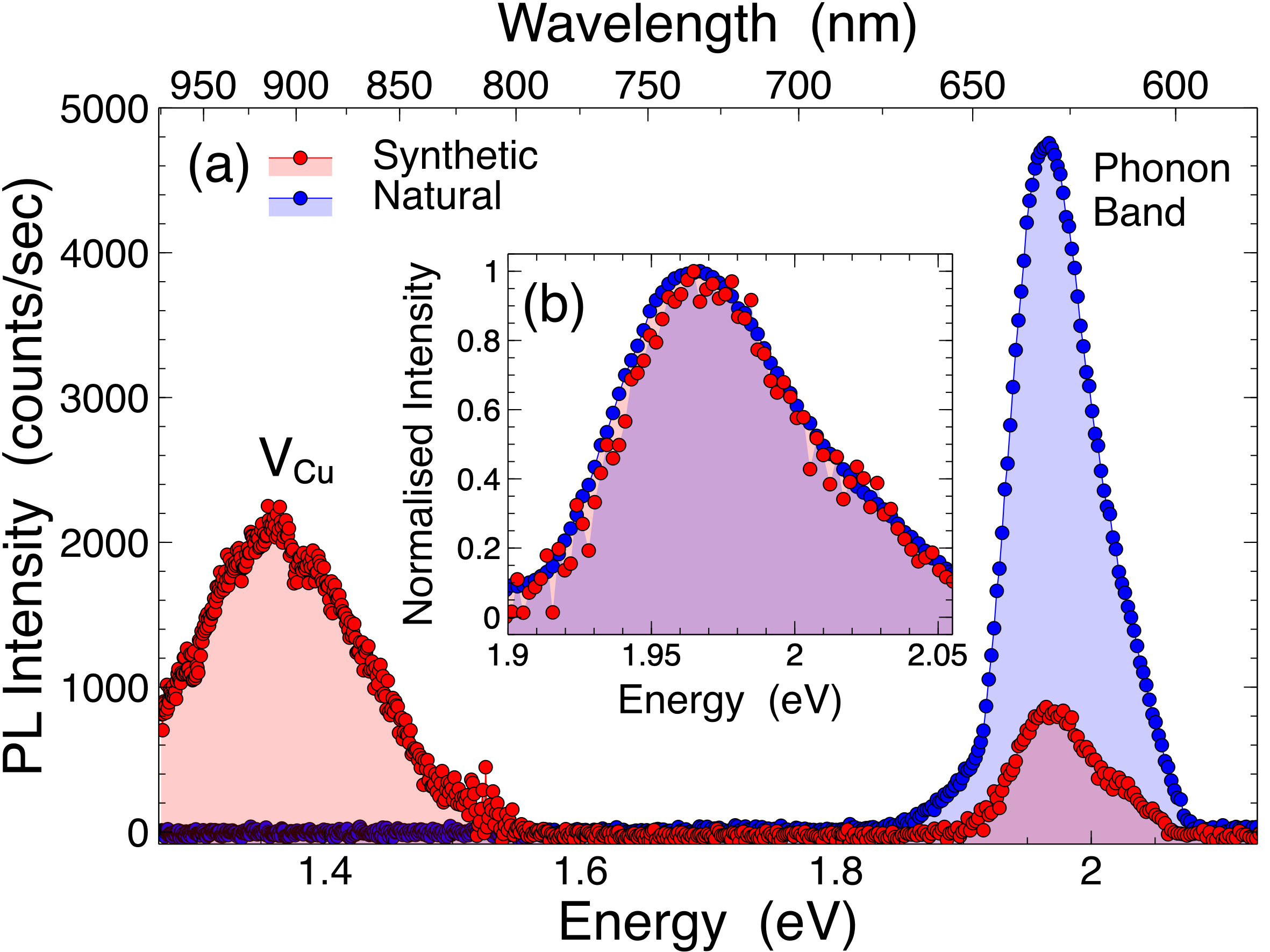}
	\caption{\label{fig:PL_RT} (a) Broad-band photoluminescence (PL) spectrum recorded at room temperature for synthetic and natural gemstone samples. A broad emission peak associated with copper vacancies (V$_{\rm Cu}$) dominates the PL emission in the synthetic sample at room temperature. Inset (b) the line-shape associated with the phonon emission band in both materials shows excellent agreement when normalized to the peak PL intensity as would be expected.}
\end{figure}

Next we turn our attention to point defects, which we study using photoluminescence (PL) and mid-infrared absorption (MIR) spectroscopy. Figure~\ref{fig:PL_RT} shows the PL spectrum measured at room temperature. Here the excitation source was a He:Ne laser operating at 543~nm. The excitation light was tightly focused onto the sample using a microscope objective with a numerical aperture of 0.6, which also served to collect the resulting luminescence. Residual excitation light was removed using a dichroic mirror, before the PL light was coupled into  multi-mode optical fiber and guided to a wide-band grating spectrometer. The signals were carefully corrected for sample tilt and the chromatic aberration of the objective, enabling a quantitative comparison of the two samples across the full spectral range.
The results are striking. For the natural sample, the dominant feature is due to quadrupole and phonon-assisted recombination of the 1S ortho-exciton. For the synthetic sample, this feature is drastically reduced, and instead the spectrum is dominated by below-gap luminescence at 1.4~eV which has previously been attributed to copper vacancies. Note that at 300~K, oxygen vacancies are expected to be ionized~\cite{koirala13}, and we do not see their luminescence at 1.70~\si{\eV} in either sample.

\begin{figure*}[!htp]
\includegraphics[width=\linewidth]{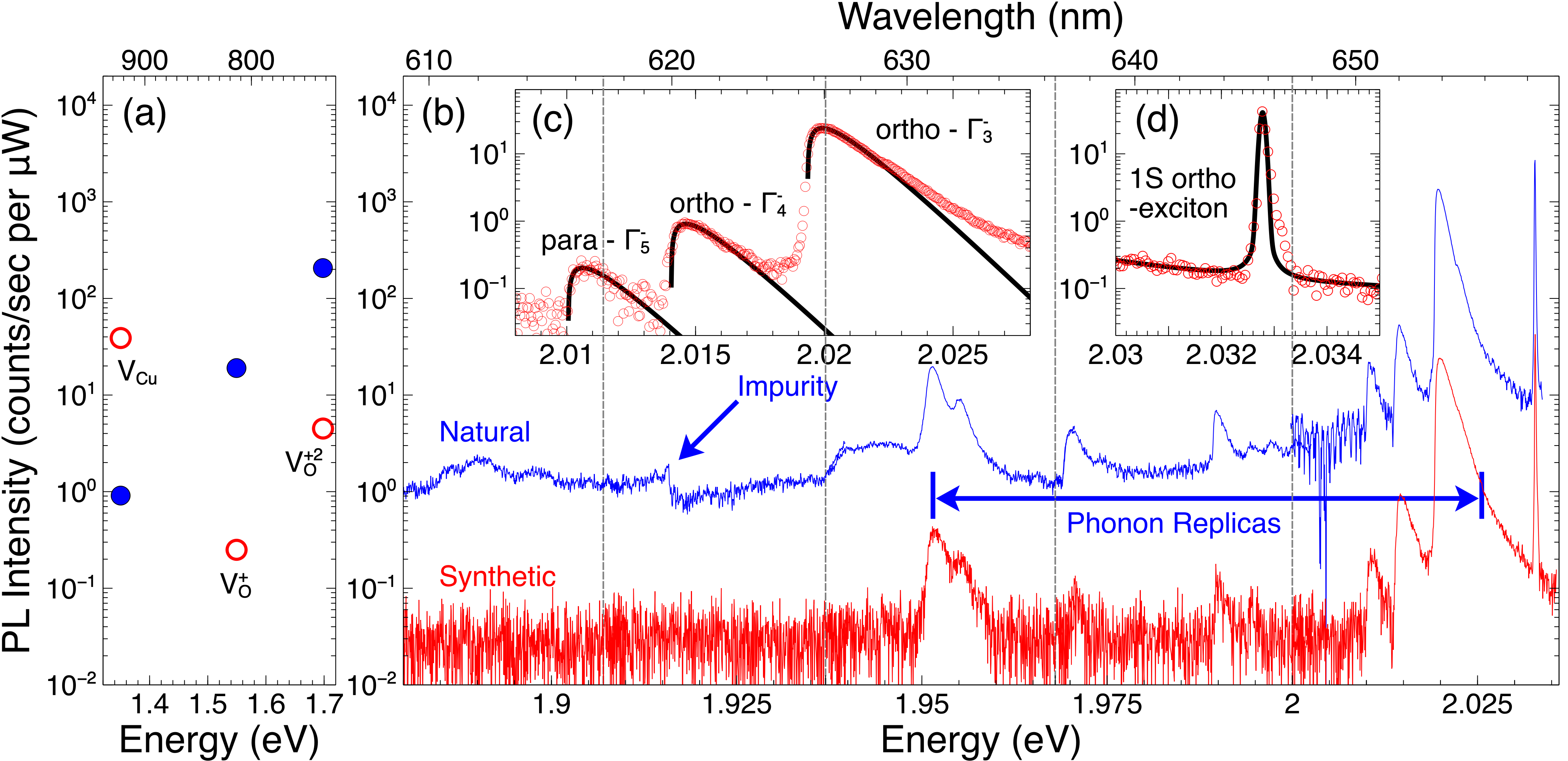} \\
\caption{High-resolution photoluminescence (PL) spectra recorded at low temperature ($\approx$~5~K) for synthetic (red) and natural gemstone (blue) Cu$_2$O normalised with respect to pump power. The subsidiary figure (a) shows the infrared PL intensity at three discrete emission wavelengths (918~nm, 800~nm, and 730~nm) associated with Cu and O vacancies, and assigned according to~\cite{frazer17}. The main figure (b) shows the visible PL in the range 609-659~nm. The insets in this figure show (c) three phonon replicas for the synthetic material on an expanded energy scale fitted (black line) to equation~\ref{eq:orthophonon}, and (d) the ortho-phonon line fitted (black line) with a pseudo-Voigt profile also on an expanded energy scale. The lines have been assigned according to~\cite{takahata18}. The photoluminescence was much brighter for the natural gemstone, which accounts for the signal-to-noise improvement. Two small peaks not associated with phonons were observed in the natural gemstone PL spectrum at 1.914~eV (647.8~nm) and 1.916~eV (647.1~nm) which we attribute to impurity luminescence.}
\label{fig:PL_characteristics}
\end{figure*}

To investigate these effects with higher resolution, we also performed PL measurement at low temperature using the same high-resolution spectrometer as in figure~\ref{fig:exciton_spectra}. For these measurements, the input slit aperture was 20~\si{\micro\meter}, corresponding to 1.16~\si{\micro\meter} at the sample plane. The corresponding spectral resolution is 70~\si{\micro}eV at 611~nm (2.01~eV) in first order.  The Cu$_2$O was illuminated by above band gap continuous-wave low-intensity radiation ($\approx 50$~\si{\micro\watt}) with wavelength \mbox{$\lambda = 532$~nm}. We performed a high-resolution study of the region near the band edge (Fig. \ref{fig:PL_characteristics}(b)), as well as localized measurements at energies associated with oxygen and copper vacancies (Fig. \ref{fig:PL_characteristics}(a)). The influence of excitation intensity on line-shape was investigated and found to be negligible. The main effect that we observed was that the temperature of the excitons increased marginally at higher excitation intensity.

The region close to the band edge shows good agreement with a previous comprehensive PL study from the Kyoto group~\cite{takahata18}. The intense sharp peak at $\sim 2.03$~eV is direct emission from the 1S ortho-exciton line. Very similar characteristics defined this spectral line in both materials. The best fit was achieved using a pseudo-Voigt profile superimposed on the exponential Bose-Einstein tail of the first phonon replica. In both materials the line exhibited a similar asymmetry in the tails of the profile. This asymmetry is highlighted for the synthetic material in the inset Fig.~\ref{fig:PL_characteristics}(d). The line fits gave a center value of 2.0328~eV for the peak and full width of 127~\si{\micro}eV for the synthetic material, and a center value of 2.0327~eV for the peak and full width of 135~\si{\micro}eV for the gemstone material. These line-widths are above our calibrated resolution limit of 70~\si{\micro}eV at 611~nm, and probably reflect the inhomogeneous broadening due to strain discussed earlier. The other large peaks in the range 1.950--2.025~eV are phonon replicas of the main 1S ortho-exciton line~\cite{takahata18}. The line-shape of these peaks can be fitted with the convolution of two terms describing the exciton density of states (DoS) and thermal occupation statistics respectively~\cite{elliott57}. The low-energy side has a rising edge, which scales with the 3D DoS, $g^{\mathrm{3D}}(E)$. The high-energy side has an exponential tail from the exciton Bose-Einstein population statistics, which can be approximated to the Boltzmann distribution $f_\mathrm{B}(E)$, since the exciton gas is of low-density. The PL intensity $I_{\mathrm{PL}}\propto g^{\mathrm{3D}}(E) f_{\mathrm{B}}(E)$ is then described by,
\begin{equation}
I_{\mathrm{PL}} = A(E - E_0)^{1/2} \exp \left[ -\dfrac{(E - E_0)}{k_\mathrm{B} T} \right] \enskip,
\label{eq:orthophonon}
\end{equation}
where $A$ is a proportionality constant and $E_0$ is the energy of the exciton at zero momentum minus the phonon energy, the latter assumed to be dispersion-less. A similar line-shape fitting function was used in~\cite{takahata18}. The fitting function gave a temperature of $\sim 13$~K and $\sim 10$~K for the exciton gases in the synthetic and natural samples respectively. This is in acceptable agreement with the nominal heat-sink temperature of 5~K, because the exciton gas is expected to be at a slightly higher temperature than the crystal lattice. Table~\ref{tab:phonons} shows the energies obtained from the fitted spectral line-shapes of several of the exciton phonon replicas, as well as the values reported in~\cite{takahata18} for comparison.
\begin{table}[htp]
\caption{\label{tab:phonons} Energy positions of the phonon assisted lines \mbox{(in~eV)}. The relative positions with respect to the resonance energy of the yellow 1S ortho-exciton are also tabulated \mbox{(in meV)} and compared with the values reported in~\cite{takahata18}. }
\begin{center}
\begin{tabular}{c>{\centering}p{2cm}>{\centering}p{2cm}c}
\hline \hline
Phonon & Center & \mbox{This work} &~\cite{takahata18} \\
& (eV) & (meV) & (meV) \\
\hline
$\Gamma_5^-$ &&  & 10.5 \\
$\Gamma_3^-$ & 2.0192 & 13.6 & 13.5 \\
$\Gamma_4^{-(1)}(\rm TO)$ & 2.0140 & 18.8 & 18.8 \\
$\Gamma_4^{-(1)}(\rm LO)$ & & & 19.3 \\
$\Gamma_2^-$ & 1.9892 & 43.6 & 43.4 \\
$\Gamma_5^+$ & 1.9693 & 63.5 & 63.2 \\
$\Gamma_4^{-(2)}(\rm TO)$ & 1.9543 & 78.5 & 78.4 \\
$\Gamma_4^{-(2)}(\rm LO)$ & 1.9507 & 82.1 & 82.0 \\
\hline
\end{tabular}
\end{center}
\label{default}
\end{table}
We have not reported the energies of the $\Gamma_5^-$ and $\Gamma_4^{-(1)}(\rm LO)$ phonon assisted transitions in this table because we are not confident that they can be reliably deconvolved from the dominant neighbouring phonon assisted transitions in our PL data. The agreement with the values reported in~\cite{takahata18} is otherwise excellent. We have, however, observed some differences between our natural gemstone PL spectrum and the data presented by the Kyoto group~\cite{takahata18}. The PL spectrum in their paper shows a sharp peak located at 1.947~eV, which sits on the red-shifted shoulder of the double-peaked phonon feature above just 1.95~eV. We see no evidence of this spectral feature in our PL spectrum. In contrast, we observe two small sharp peaks at 1.914~eV and 1.916~eV, which are not present in~\cite{takahata18}. Unfortunately, the signal-to-noise ratio for our synthetic Cu$_2$O PL spectrum is insufficient to resolve these spectral features. We speculate that the origin of these peaks is due to different trace impurities to those found in the Kyoto samples. We have also recorded the intensity of the PL emission at three discrete wavelengths known to be associated with copper and oxygen vacancies~\cite{frazer15,frazer17} in both synthetic and natural samples (Fig.~\ref{fig:PL_characteristics}(a)). Measurements at 1.35~eV (918~nm) confirm the observation of strong luminescence from copper vacancies observed in the room temperature spectrum. Measurements at 1.70~eV (800~nm) and 1.55~eV (730~nm) and are associated with singly and doubly ionized oxygen vacancies, V$_{\rm O}^{+1}$ and V$_{\rm O}^{+2}$, respectively~\cite{frazer17}. These peaks are not observed in the room temperature data. Here we find that the luminescence from these states is strongest in the natural sample.

In summary, the region close to the band edge indicates a narrow 1S exciton and all the expected phonon modes in both samples. Apart from the stronger overall luminescence of the natural sample at low temperature, the spectrum of the two samples is largely identical in this region. However, the PL measurements indicate a large excess of copper vacancies in the synthetic material, which is presumably the result of the fabrication process (oxidation of the copper rod).

\subsection{Mid-Infrared Spectroscopy}
To search for other impurities that might alter the Rydberg spectrum, we also performed MIR measurements to look for shallow bound states such as those observed in silicon~\cite{jagannath81}. Much like excitons, impurity states are good sensors of the local crystal environment around the defect, and their inhomogeneous line-width is another indicator of the material quality~\cite{lynch10}. The spectra were recorded using a Bruker Vertex 80V Fourier transform spectrometer (FTS). The samples were mounted on the cold-finger of a helium flow cryostat located in the spectrometer absorption compartment, and the optical path within the spectrometer was fully evacuated. Sample illumination was provided by the internal FTS globar, and the transmitted interferogram was recorded on a liquid nitrogen cooled mercury cadmium telluride (MCT) detector. Cooling was provided by a closed-cycle ColdEdge\textsuperscript{\texttrademark} ``Stinger'' helium re-circulator.

\begin{figure}[htp]
\includegraphics[width=8.3cm]{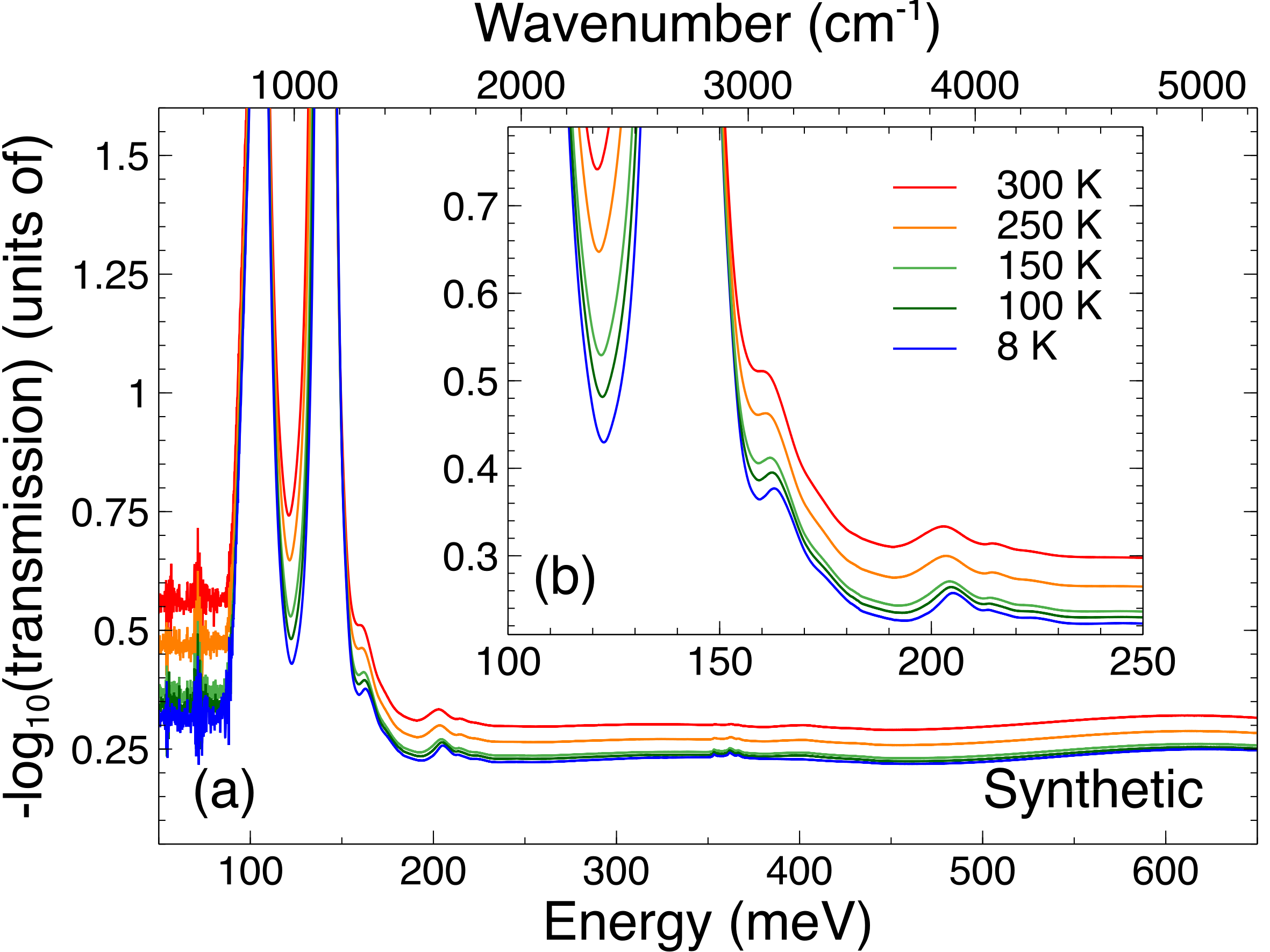} \\
\vspace{5mm}
\includegraphics[width=8.3cm]{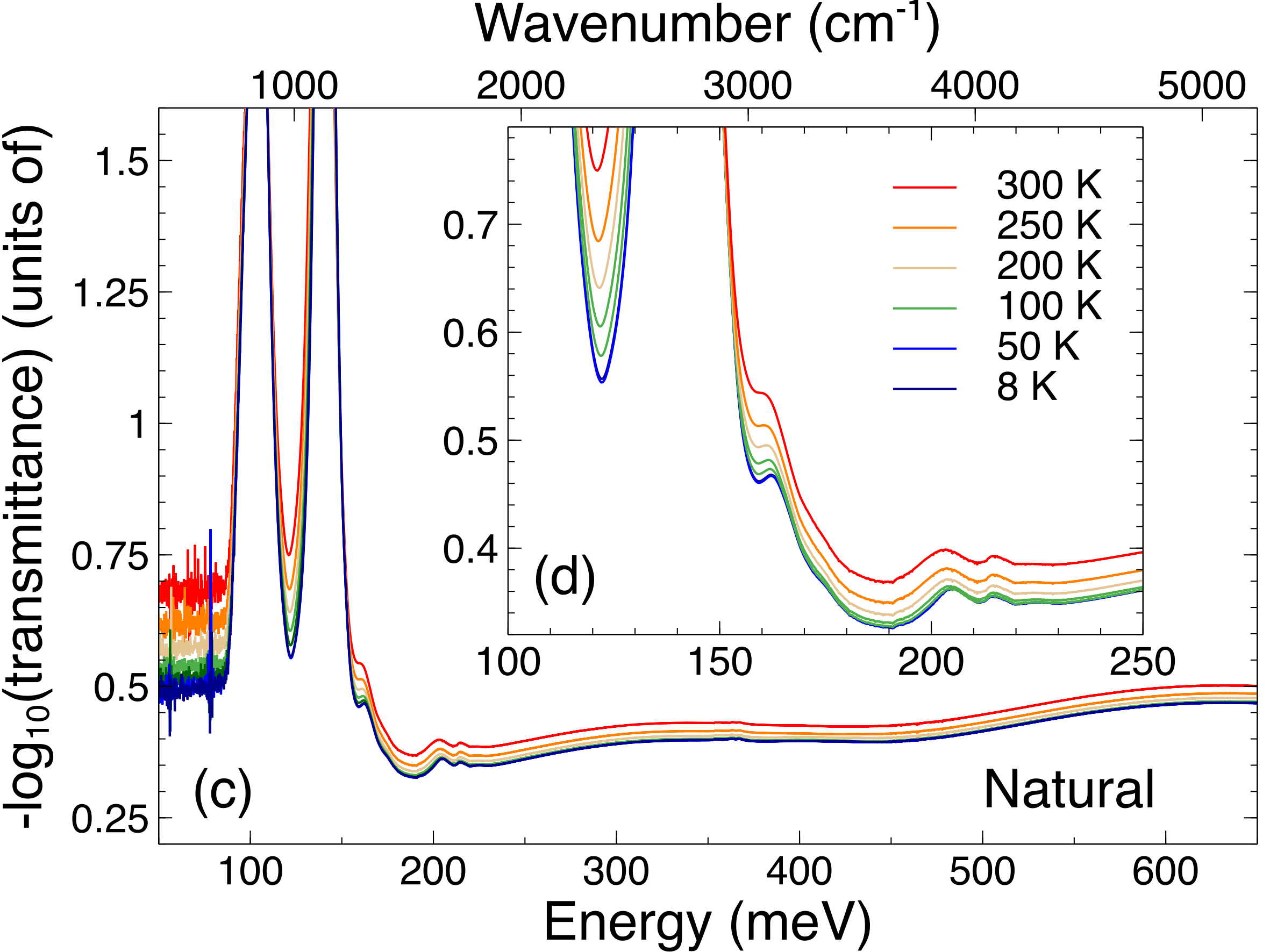} \\
\caption{(a) MIR transmission spectra of a 1 mm thick slice of synthetic Cu$_2$O crystal recorded at various temperatures between 8~K and 300~K. The increased noise below 80~meV indicates where the MCT detector response falls off at long wavelengths. The inset (b) shows the long-wavelength region near the Reststrahlen band edge on an expanded energy scale. (c) MIR transmission spectra of a 1 mm thick slice of natural Cu$_2$O gemstone acquired at a location where \mbox{high-$n$} excitons were observed and measured under similar conditions. The inset (d) similarly shows the region near the Reststrahlen band edge on an expanded energy scale.}
\label{fig:MIR_spectra}
\end{figure}

Figure \ref{fig:MIR_spectra} shows the mid-infrared transmission spectrum for both (a) synthetic and (c) natural samples recorded for a range of temperatures between 8~K and 300~K. The samples exhibited very similar spectra with only some minor differences. The spectra are dominated by two Reststrahlen bands between 95-115~meV and 135-155~meV. Some additional peaks become more visible when the y-axis is expanded as shown in the inset. The shoulder at 164~meV relates to the second Reststrahlen band and is present in both samples. In both cases this shoulder sharpens to a discernible peak at lower temperatures. We attribute this peak to the highest energy longitudinal optical (LO) phonon. The peak at 205~meV is also present in both samples. This peak shows a weak temperature dependence, becoming slightly sharper at lower temperature. While also present in both samples, a second peak at 215~meV appears stronger in the natural sample. The physical origin of these spectral features is open to debate. It is generally accepted that synthetically grown Cu$_2$O is always p-type~\cite{scanlon10}, and acceptor levels due to copper vacancies lying at 220~meV and 470~meV have been proposed. Previous density functional theory (DFT) calculations performed by the same research group~\cite{scanlon09} pointed towards a hole trap 450~meV above the valence band edge attributed to single copper vacancies, and a second hole trap around 250~meV tentatively attributed to a copper di-vacancies. We do not observe the 450~meV feature in either sample but the peaks we observe at 205~meV and 215~meV could fit either the copper vacancy or di-vacancy prediction. There is also some experimental evidence to support this hypothesis~\cite{paul06}. Two hole trap levels consistent with this description have been observed in the deep level transient spectrum (DLTS) of polycrystalline p-type Cu$_2$O epilayers grown on top of intrinsic zinc oxide (i-ZnO) and fabricated into mesa heterostructures. Theoretical calculations exploiting the generalized gradient approximation to DFT have also been used to predict a single copper vacancy level at 280~meV in p-type Cu$_2$O~\cite{raebiger07}. Another possibility is that this feature could be associated with multi-phonon absorption. The spectral location matches twice the energy of the observed phonon bands, and the fact that the spectral feature is observed in two samples of completely different origin strongly suggests that it is intrinsic to the material.

\section{Conclusions}

In summary, there are strong similarities between the excitonic spectrum that we observe in the synthetic material and that observed in high-quality natural gemstones. The principal differences are a cut-off at lower principal quantum number, the emergence of additional peaks, and an overall shift in the spectrum. We note however that the observation of $n=10$ in synthetic material is comparable to the state-of-the-art~\cite{ito98}. The large spatial extent of the \mbox{high-$n$} exciton wave-functions means that they are sensitive to lattice imperfections. For example, using the formula for the Bohr radius of the P-exciton series quoted in~\cite{kazimierczuk14}, together with \mbox{$a_\mathrm{B} = 1.11$~nm}, would give $\langle r_{10} \rangle = 165$~nm, which corresponds to 386 lattice constants. Measurements of the 1S ortho-exciton width and the spatial variation of the Rydberg spectrum suggest that inhomogeneous strain, while present, cannot explain the origin of these features.  PL measurements demonstrate a large excess of copper vacancies as well as the quenching of the band-edge luminescence associated with intrinsic effects in the synthetic sample. A search for additional shallow donor states in the MIR region did not produce a clear result. The hypothesis proposed in~\cite{agekyan77}, and discussed earlier in this work, was that point defects lead to the main line broadening of the exciton line, while large-size defects (dislocations, for example) originate a low-gradient electric field which induces forbidden lines. Broadly speaking, our observations of a modified Rydberg spectrum and a large excess of copper vacancies is consistent with this interpretation. We have not attempted to address these problems yet, but there are many promising post-growth treatments to try. For example, annealing at high-temperature has been successfully employed to remove the CuO-walled voids~\cite{schmidt_whitley74, ito98} and reduce the concentration of copper vacancies~\cite{kaufman86}. The mechanical polishing procedure employed to produce samples for optical experiments may itself introduce problems that are specific to the sample origin. In the synthetic material, mechanical polishing can open some of the CuO-walled voids and this can deposit unspecified material into the polishing slurry. We have on occasion observed the appearance of new scratches and micro-cracking of the surface during the polishing process when this happens. However, there are fabrication strategies such as chemical etching~\cite{takahata18} or ion-beam milling that might mitigate the consequences of this type of surface damage.

Our results are important because they demonstrate that it is possible to produce large scale bulk synthetic material with comparable optical properties to natural gemstone material. This represents a significant step forward because it shows that there is a viable technological route towards harnessing the quantum optical properties of this extraordinary semiconductor material.

\begin{acknowledgments}
This work was supported by the Engineering and Physical Sciences Research Council (EPSRC), United Kingdom, through research grants EP/P011470/1 and EP/P012000/1. The authors also acknowledge seedcorn funding from Durham University. LG acknowledges financial support from the UK Defence and Scientific Technology Laboratory via an EPSRC Industrial Case award. SM and RPS would like to thank the Royal Society Yusuf Hamied International Exchange Award(IES/R2/181048) which has made this collaboration possible. We are grateful to Ian Chaplin and Sophie Edwards (Durham University, Department of Earth Sciences) for the slicing and polishing of some of the samples used in this work.  Information on the data underpinning the results presented here, including how to access them, can be found in the Cardiff University data catalogue at [DOI].
\end{acknowledgments}

\bibliography{references_oct2020mod}

\end{document}